\documentclass[sigplan,10pt]{acmart}
\renewcommand\footnotetextcopyrightpermission[1]{}
\usepackage[utf8]{inputenc}
\usepackage{totcount}
\usepackage{listings}
\usepackage[T1]{fontenc}
\usepackage{enumitem}

\usepackage{algorithm}
\usepackage[noend]{algpseudocode}
\algrenewcommand\algorithmicrequire{\textbf{Input:}}
\algrenewcommand\algorithmicensure{\textbf{Output:}}

\newtheorem{definition}{Definition}
\newtheorem{theorem}{Theorem}

\newtheorem{corollary}{Corollary}[theorem]

\newtoggle{shownotes}
\toggletrue{shownotes}
\newtotcounter{numnotes}

\newcommand{\system}{\textsc{aft}}
\newcommand{\awb}{Atomic Write Buffer}

\newcommand{\figwidth}{.48\textwidth}

\newcommand{\newtext}[1]{\textcolor{black}{#1}}

\newtoggle{showtodos}
\toggletrue{showtodos}
\newtotcounter{numtodos}

\newcommand{\checkforcrud}{
  \ifboolexpr{test{\ifnumcomp{\totvalue{numnotes}}{>}{0}} or
              test{\ifnumcomp{\totvalue{numtodos}}{>}{0}}}{
    \pagecolor{red!20}
  }{}
}

\newcommand{\smallitem}[1]{\vspace{0.3em}\noindent\textbf{#1}}
\newcommand{\smallitembot}{\vspace{0.5em}\noindent}

\acmConference[]{}{}{}

\renewcommand\footnotetextcopyrightpermission[1]{} 

\begin{document}

\date{}

\title{A Fault-Tolerance Shim for Serverless Computing}

\author{Vikram Sreekanti}
\affiliation{UC Berkeley}

\author{Chenggang Wu}
\affiliation{UC Berkeley}

\author{Saurav Chhatrapati}
\affiliation{UC Berkeley}

\author{Joseph E. Gonzalez}
\affiliation{UC Berkeley}

\author{Joseph M. Hellerstein}
\affiliation{UC Berkeley}

\author{Jose M. Faleiro}
\affiliation{Microsoft Research}

\renewcommand{\shortauthors}{Sreekanti et al.}

\begin{abstract}

Serverless computing has grown in popularity in recent years, with an increasing number of applications being built on Functions-as-a-Service (FaaS) platforms.
By default, FaaS platforms support retry-based fault tolerance, but this is insufficient for programs that modify shared state, as they can unwittingly persist partial sets of updates in case of failures.
To address this challenge, we would like atomic visibility of the updates made by a FaaS application. 

In this paper, we present \system{}, an atomic fault tolerance shim for serverless applications.
\system{} interposes between a commodity FaaS platform and storage engine and ensures atomic visibility of updates by enforcing the \emph{read atomic} isolation guarantee.
\system{} supports new protocols to guarantee read atomic isolation in the serverless setting.
We demonstrate that \system{} introduces minimal overhead relative to existing storage engines and scales smoothly to thousands of requests per second, while preventing a significant number of consistency anomalies.

\end{abstract}

\checkforcrud{}

\settopmatter{printfolios=true}
\maketitle
\thispagestyle{empty}

\section{Introduction} \label{sec:intro}

Serverless computing is an emerging area of focus in research~\cite{excamera, pywren, akkus2018sand, bv-serverless, serverless-cidr19, pocket, baldini2017serverless, van2017spec, gan2019open, fouladi2019laptop} and industry~\cite{awscasestudies}.
Its key attraction is simplicity: Users upload code and select trigger events (e.g., API invocation, file upload), and the cloud provider transparently manages deployment, scaling, and billing for those programs.
Users are charged for resource time \emph{used} rather than having to plan and provision resources in advance.

In today's public clouds, serverless computing most often refers to Functions-as-a-Service (FaaS).
FaaS platforms allow users to construct applications in high-level languages while imposing limitations on what those applications can do.
One key limitation is the requirement that programs be stateless---requests are not guaranteed to be routed to any particular instance of a program, so developers cannot rely on machine-local state to handle requests.
As a result, applications built on FaaS infrastructure must either be purely functional or modify state in a shared storage system like Amazon Web Services' S3 or Google Cloud's Datastore.

FaaS platforms provide some measure of fault tolerance with retries---systems like AWS Lambda and Azure Functions automatically retry functions if they fail, and clients typically will re-issue requests after a timeout.
Retries ensure that functions are executed \emph{at-least once}, and cloud providers encourage developers to write idempotent programs~\cite{lambda-best-practices} because idempotence logically ensures \emph{at-most once} execution.
Combining retry-based at-least once execution and idempotent at-most once execution would seem to guarantee \emph{exactly once} execution,  a standard litmus test for correct fault handling in distributed systems.

This idempotence requirement is both unreasonable for programmers---as programs with side-effects are ubiquitous---and also insufficient to guarantee exactly once serverless execution.
To see why idempotence is insufficient, consider a function $f$ that writes two keys, $k$ and $l$, to storage.
If $f$ fails between its writes of $k$ and $l$, parallel requests might read the new version of $k$ while reading an old version of $l$.
Even if $f$ is idempotent---meaning that a retry of $f$ would write the same versions of $k$ and $l$---the application has exposed a \emph{fractional} execution in which some updates are visible and others are not.
As a result, developers are forced to explicitly reason about the \emph{correctness} of their reads in addition to the idempotence of their applications---an already difficult task.

We propose that in a retry-based fault-tolerance model, \emph{atomicity} is necessary to solve these challenges: Either all the updates made by an application should be visible or none of them should.
Returning to the example above, if $f$ fails between the writes of $k$ and $l$, an atomicity guarantee would ensure that $f$'s version of $k$ would not be visible to other functions without $f$'s version of $l$.
Atomicity thus prevents fractional executions from becoming visible.

A simple solution here is to use serializable transactions.
Functions have well-defined beginnings and endings, making a transactional model a natural fit to guarantee atomicity for FaaS platforms.
In reality, each logical request in an application will likely span multiple functions---as it would be difficult to pack a whole application into one function---meaning transactions should span \emph{compositions} of functions.
However, strongly consistent transactional systems have well-known scaling bottlenecks~\cite{brewercap,chandra2007paxos} that make them ill-suited for serverless settings, which must accommodate unbounded numbers of client arrivals.

Instead, we turn to \emph{read atomicity}, a coordination-free isolation level introduced in \cite{bailis2014ramp}.
Intuitively, read atomic isolation guarantees that transactions do not leak partial side effects to each other, which is the requirement described above.
However, the original implementation of read atomic isolation---termed RAMP in~\cite{bailis2014ramp}---makes key limiting assumptions that are unreasonable in our setting.
First, their storage system forbids replication, which limits locality and scalability of reads. 
Second, they assume that every transaction's read and write sets are predeclared, which is unreasonable for interactive applications that make dynamic decisions.

\subsection{A Fault-Tolerance Shim}

Our goal in this paper is to provide fault tolerance in the context of widely-used FaaS platforms and storage systems.
To that end, we present \system{}, an Atomic Fault Tolerance shim for serverless computing.
\system{} provides fault-tolerance for FaaS applications by interposing between a FaaS platform (e.g., AWS Lambda, Azure Functions) and a cloud storage engine (e.g., AWS S3, Google Cloud BigTable).
Updates written to storage during a single logical request---which may span multiple functions---are buffered by \system{} and atomically committed at request end.
\system{} enforces the read atomic isolation guarantee, ensuring that transactions never see partial side effects and only read data from sets of atomic transactions.
Given the unsuitability of the original RAMP protocols for serverless infrastructure, we develop new protocols to support read atomic isolation over shared storage backends.
Importantly, \system{} maintains a high measure of flexibility by only relying on the storage engine for durability.

The contributions of this paper are the following:
 
\begin{itemize}[leftmargin=*]
    \item The design of \system{}, a low-overhead, transparent fault tolerance shim for serverless applications that is flexible enough to work with many combinations of commodity compute platforms and storage engines.
    
    \item A new set of protocols to guarantee read atomic isolation for shared, replicated storage systems.
    
    \item A garbage collection scheme for our protocols that significantly reduces the storage overheads of read atomic isolation. 
    
    \item A detailed evaluation of \system{}, demonstrating that it imposes low latency penalties and scales smoothly to hundreds of clients and thousands of requests per second, while preventing consistency anomalies.
    
\end{itemize}
\section{Background and Motivation} \label{sec:background}

In this section, we describe prior work on read atomicity and its guarantees ($\S$\ref{sec:bg-ra}), and we explain the technical challenges in providing read atomicity for serverless applications ($\S$\ref{sec:bg-assumptions}).

\subsection{Read Atomic Isolation} \label{sec:bg-ra}

The read atomic isolation guarantee, introduced by Bailis et al. in~\cite{bailis2014ramp}, aims to ensure that transactions do not view partial effects of other transactions.
Bailis et al. provide the following definition: ``A system provides Read Atomic isolation (RA) if it prevents fractured reads anomalies and also prevents transactions from reading uncommitted, aborted, or intermediate data.''
In this paper, we refer to reads of ``uncommitted, aborted, or intermediate'' data as \emph{dirty reads}.
A \emph{fractured read} happens when, ``... transaction $T_i$ writes versions $x_m$ and $y_n$ (in any order, with $x$ possibly but not necessarily equal to $y$), [and] $T_j$ [later] reads version $x_m$ and version $y_k$, and $k < n$.''
The read atomic isolation level is a good fit for the serverless setting because it enforces atomic visibility of updates without strong consistency or coordination.

\subsection{Challenges and Motivation} \label{sec:bg-assumptions}

The protocols introduced in \cite{bailis2014ramp} make two assumptions that are unreasonable for serverless applications: pre-declared read/write sets and a linearizable, unreplicated, and sharded storage backend.
Relaxing these assumptions enables us to bring read atomic isolation to the serverless setting but raises new challenges around consistency and visibility.

\smallitem{Read and Write Sets}.
\cite{bailis2014ramp} requires that each transaction declares its read and write sets in advance, in order to correctly ensure that the transaction's operations adhere to the definition of read atomicity above.
We relax this assumption, allowing requests to issue reads and writes without restriction, and we develop new read atomic protocols that allow us to dynamically construct atomic read sets ($\S$\ref{sec:atomicity}).
The drawback of this flexibility is that clients may be forced to read data that is more stale than they would have under the original RAMP protocol, and in rare cases, a request may be forced to abort because there are no valid key versions for it to read.
We explain this tradeoff in more detail in $\S$\ref{sec:atomicity-staleness}.

\smallitem{Shared Storage Backends}.
The RAMP protocols assume linearizable, unreplicated, and shared-nothing storage shards, each of which operate independently.
Each shard is the sole ``source of truth'' for the set of keys it stores, which can lead to scaling challenges in skewed workloads.
This design is incompatible with standard cloud applications, where workloads can often be highly skewed~\cite{Taft:2014:EFE:2735508.2735514, beame2014skew} and require strong durability.
To that end, \system{} offers read atomic isolation in a shim layer over a durable shared storage backend \newtext{without requiring the storage layer to provide consistency guarantees and partitioning}.
We avoid the scaling pitfalls of RAMP by letting all nodes commit data for all keys.
This requires carefully designing commit protocols for individual nodes ($\S$\ref{sec:atomicity-writes}) and ensuring that nodes are aware of transactions committed by their peers ($\S$\ref{sec:distributed}).
Our coordination-free and fungible node design leads to a potential ballooning of both data and metadata, which we address in $\S$\ref{sec:gc}.

\smallitem{Serverless Applications}.
Serverless applications are typically compositions of multiple functions.
We model each request as a linear composition of one or more functions executing on a FaaS platform.
\system{} must ensure atomic reads and writes across all functions in the composition, each of which might be executed on a different machine---this informs our design.
In effect, we must support a ``distributed'' client session as the transaction moves across functions, and we must ensure that retries upon failure guarantee idempotence.
We discuss these issues in more detail in $\S$\ref{sec:atomicity}.
\section{Achieving Atomicity} \label{sec:atomicity}
In this section, we describe how \system{} achieves atomic writes and reads at a single node.
We first discuss \system{}'s API, architecture, and components ($\S$\ref{sec:atomicity-arch}).
We then formally state the guarantees we make ($\S$\ref{sec:atomicity-guarantees}); we describe \system{}'s protocol for each of the guarantees in $\S$\ref{sec:atomicity-writes}-$\S$\ref{sec:atomicity-rr-ryw}. 
In $\S$\ref{sec:distributed}, we discuss how\system{} operates in distributed settings.

\subsection{Architecture and API} \label{sec:atomicity-arch}

\system{} offers transactional key-value store API, shown in Table~\ref{table:api}.
We refer to each logical request (which might span multiple FaaS functions) as a transaction.
A new transaction begins when a client calls \texttt{StartTransaction}, and the transaction is assigned a globally-unique UUID.
At commit time, each transaction is given a commit timestamp based on the machine's local system clock.
In the rest of the paper, we refer to this $\langle timestamp ,uuid\rangle$ pair as the transaction's ID.
\system{} uses each transaction's ID to ensure that its updates are only persisted once, assuring idempotence in the face of retries---as discussed in $\S$\ref{sec:intro}, guaranteeing idempotence and atomicity results in exactly once execution semantics.
We do not rely on clock synchronization for the correctness of our protocols, and we only use system time to ensure relative freshness of reads.
As a result, we need not coordinate to ensure timestamp uniqueness---ties are broken by lexicographically comparing transactions' UUIDs.

\begin{figure}[t]
  \centering
    \includegraphics[width=\figwidth]{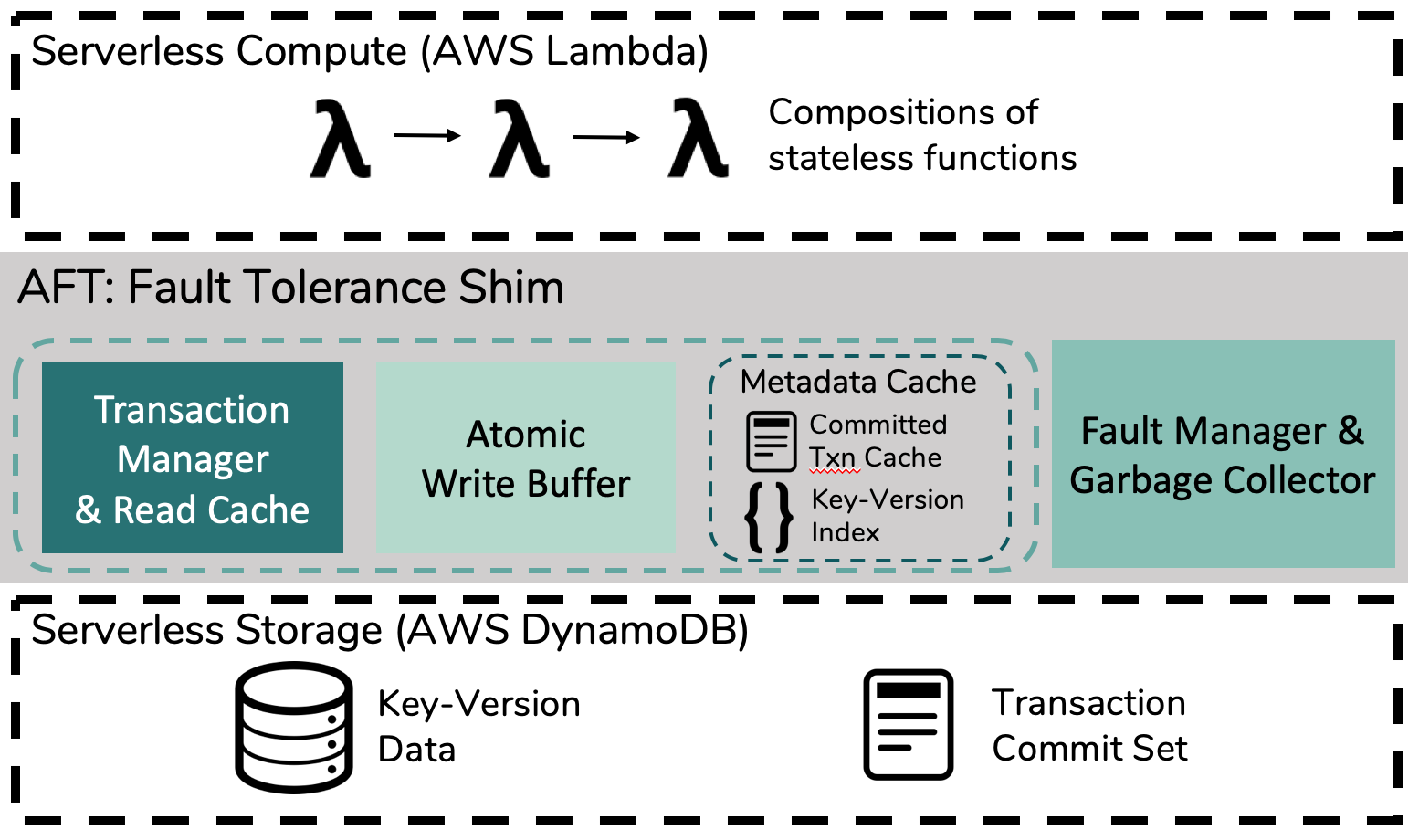}
  \caption{\small
    A high-level overview of the \system{} shim in context.
  }
  \label{fig:arch}
  \vspace{-0.5em}
\end{figure}

\begin{table}[t]
\footnotesize
\begin{tabular}{|p{.4\columnwidth}|p{.5\columnwidth}|}
\hline
\textbf{API} & \textbf{Description} \\
\hline
\textbf{\texttt{StartTransaction()->txid}} & Begins a new transaction and returns a transaction ID. \\
\hline
\textbf{\texttt{Get(txid, key)->value}} & Retrieves \texttt{key} in the context of the transaction keyed by \texttt{txid}. \\
\hline
\textbf{\texttt{Put(txid, key, value)}} & Performs an update for transaction \texttt{txid}. \\
\hline
\textbf{\texttt{AbortTransaction(txid)}} & Aborts transaction \texttt{txid} and discards any updates made by it. \\
\hline
\textbf{\texttt{CommitTransaction(txid)}} & Commits transcation \texttt{txid} and persists its updates; only acknowledges after all data and metadata has been persisted. \\
\hline
\end{tabular}
\caption{\small
    \system{} offers a simple transactional key-value store API.
    All \texttt{get} and \texttt{put} operations are keyed by the ID transaction within which they are executing.
}
\label{table:api}
\end{table}

Each transaction sends all operations to a single \system{} node.
Within the bounds of a transaction, clients interact with \system{} like a regular key-value store, by calling \texttt{Get(txid, key)} and \texttt{put(txid, key, value)}.
When a client calls \texttt{CommitTransaction}, \system{} assigns a commit timestamp, persists all of the transaction's updates, and only acknowledges the request once the updates are durable.
At any point during its execution, if a client calls \texttt{AbortTransaction}, none of its updates are made visible, and the data is deleted.

Figure~\ref{fig:arch} shows a high-level overview of the system architecture.
Each \system{} node is composed of a transaction manager, an atomic write buffer, and a local metadata cache. 
The atomic write buffer gathers each transaction's writes and is responsible for atomically persisting them at commit time.
The transaction manager tracks which key versions each transaction has read thus far and is responsible for enforcing the read atomicity guarantee described later in this section.
\system{} maintains a Transaction Commit Set  storage, which holds the ID of each committed transaction and its corresponding write set.
We defer discussion of fault management and garbage collection to $\S$\ref{sec:distributed} and $\S$\ref{sec:gc}, respectively.

Algorithm~\ref{alg:read} ($\S$\ref{sec:atomicity-reads}) requires access to the list of recently committed transactions.
To avoid metadata fetches on each read, \system{} caches the IDs of recently committed transactions and locally maintains an index that maps from each key to the recently created versions of that key.
When an \system{} node starts (e.g., after recovering from failure), it bootstraps itself by reading the latest records in the Transaction Commit Set to warm its metadata cache.
We discuss how the metadata cache is pruned by the garbage collection processes in $\S$\ref{sec:gc}.

In addition to a metadata cache, \system{} has a data cache, which stores values for a subset of the key versions in the metadata cache.
The data cache improves performance by avoiding storage lookups for frequently accessed versions; we measure its effects in $\S$\ref{sec:eval-caching}.

\newtext{\system{} make no assumptions about the serverless compute layer it is interacting with---it simply responds to the API calls described in Table~\ref{table:api}.
The only assumption \system{} makes about the underlying storage engine is that updates are durable once acknowledged.
\system{} does not rely on the storage engine to enforce any consistency guarantees or to immediately make keys visible after they are written.}

\subsection{Definitions} \label{sec:atomicity-guarantees}

A transaction $T$'s ID is denoted by its subscript: transaction $T_i$ has ID $i$.
We say that $T_i$ is newer than $T_j$ ($T_i > T_j$) if $i > j$---we describe how IDs are compared in $\S$\ref{sec:atomicity-arch}.
A key without a subscript, $k$, refers to any version of that key; $k_i$ is the version of key $k$ written by transaction $T_i$.
Each key has a \texttt{NULL} version, followed by many non-null versions.
Importantly, key versions are hidden from users: Clients requests reads and writes of keys, and \system{} automatically determines which versions are compatible with each request.

We define a transaction $T_i$'s write set, $T_i.writeset$, as the set of key versions written by this transaction.
Similarly, each key version has a cowritten set: If $T_i$ writes $k_i$, $k_i.cowritten$ is the same as $T_i.writeset$.

As described in $\S$\ref{sec:bg-ra}, read atomic isolation requires preventing \emph{dirty reads} and preventing \emph{fractured reads}.
\system{} ensures two simple properties in addition to read atomicity that most programmers are familiar with: \emph{read-your-writes} and \emph{repeatable read}.
We describe these four properties briefly.

\smallitem{Dirty Reads}.
To prevent dirty reads, \system{} guarantees that if transaction $T_i$ reads key version $k_j$ written by transaction $T_j$, $T_j$ must have successfully committed.

\smallitem{Fractured Reads}.
To avoid fractured reads, each transaction's read set must form an \textit{Atomic Readset} defined below:

\begin{definition}[Atomic Readset] \label{def:atomic-readset}

Let $R$ be a set of key versions.
$R$ is an \emph{Atomic Readset} if $\forall k_i \in R, \forall l_i \in k_i.cowritten, l_j \in R \Rightarrow j \geq i$.

\end{definition}

In other words, for each key version $k_i$ read by transaction $T_j$, if $T_j$ also reads a version of key $l$ that was cowritten with $k_i$ (i.e., $T_j$ reads $l_i$), we return a version of $l$ that is no older than $l_i$.
\newtext{Consider a scenario where there are two committed transactions in storage, $T_1: \{ l_1 \}$ and $T_2: \{ k_2, l_2 \}$, where the set following the colon denotes the set of key versions written by the corresponding transaction.
If a new transaction, $T_n$ first requests $k$ and reads $k_2$, a subsequent read of $l$ must return a version of $l$ that is at least as new as $l_2$ (i.e., $l_2$ or some newer version).
If $T_n$ were to read $l_1$, that would violate Definition~\ref{def:atomic-readset} because $l$ is in the cowritten set of $k_2$, and $l_1 < l_2$.}

\smallitem{Read Your Writes}. 
Guaranteeing read your writes requires ensuring that a transaction reads the most recent version of a key it previously wrote.
Say a transaction $T_i$ writes key version $k_{i_1}$; a following read of $k$ should return $k_{i_1}$.
If $T_i$ then proceeds to write $k_{i_2}$, future reads will return $k_{i_2}$.

\smallitem{Repeatable Read}.
Repeatable read means that a transaction should view the same key version if it requests the same key repeatedly: If $T_i$ reads version $k_j$ then later requests a read of $k$ again, it should read $k_j$, \emph{unless} it wrote its own version of $k$, $k_i$, in the interim.
Note that this means the read-your-writes guarantee will be enforced at the expense of repeatable reads.

\subsection{Preventing Dirty Reads} \label{sec:atomicity-writes}

\system{} implements atomic updates via a simple write-ordering protocol. 
\system{}'s \awb{} sequesters all the updates for each transaction.
When \texttt{CommitTransaction} is called, \system{} first writes the transaction's updates to storage.
Once all updates have successfully been persisted, \system{} writes the transaction's write set, timestamp, and UUID to the Transaction Commit Set in storage.
Only after the Commit Set is updated does \system{} acknowledge the transaction as committed to the client and make the transaction's data visible to other requests.
If a client calls \texttt{AbortTransaction}, its updates are simply deleted from the \awb, and no state is persisted in the storage engine.
This protocol is carefully ordered to ensure that dirty data is never visible.

Crucially, to avoid coordination, \system{} does not overwrite keys in place: Each key version is mapped to a unique storage key, determined by its transaction's ID. 
This naturally increases \system{}'s storage and metadata footprints; we return to this point in $\S$\ref{sec:gc}, where we discuss garbage collection.

\newtext{
\system{} prevents dirty reads by only making a transaction's updates visible to other transactions \emph{after} the correct metadata has been persisted.
As we will see in Section~\ref{sec:atomicity-reads}, \system{} only allows reads from transactions that have already committed by consulting the local committed transaction cache (see Section~\ref{sec:atomicity-arch}).
}

\newtext{
For long-running transactions with large update sets, users might worry that forcing all updates to a single \awb{} means that the update set must fit in memory on a single machine.
However, when an \awb{} is saturated, it can proactively write intermediary data to storage.
The protocols described in this section guarantee that this data will not be made visible until the corresponding commit record is persisted.
If a \system{} node fails after such a write happens but before a commit record is written, the intermediary writes must be garbage collected; we discuss this in Section~\ref{sec:gc}.
}

\subsubsection{Atomic Fault Tolerance} \label{sec:atomic-write-ft}

This write-ordering protocol is sufficient to guarantee fault tolerance in the face of application and \system{} failures.
When an application function fails, none of its updates will be persisted, and its transaction will be aborted after a timeout.
If the function is retried, it can use the same transaction ID to continue the transaction, or it can begin a new transaction.

There are two cases for \system{} failures.
If a transaction had not finished when an \system{} node fails, we consider its updates lost, and clients must redo the whole transaction.
If the transaction had finished and called \texttt{CommitTransaction}, we will read its commit metadata (if it exists) during the \system{} startup process ($\S$\ref{sec:atomicity-arch}). 
If we find commit metadata, our write-ordering protocol ensures that the transaction's key versions are persisted; we can declare the transaction successful.
If no commit record is found, the client must retry.

\subsection{Preventing Fractured Reads} \label{sec:atomicity-reads}

In this section, we introduce \system{}'s atomic read protocol.
We guarantee that after every consecutive read, the set of key versions read thus far forms an Atomic Readset (Definition~\ref{def:atomic-readset}).
Unlike in \cite{bailis2014ramp}, we do not require that applications declare their read sets in advance.
We enable this flexibility via versioning---the client may read older versions because prior reads were from an earlier timestamp.

\newtext{
\system{} uses the metadata mentioned in Section~\ref{sec:atomicity-arch} to ensure that reads are correct. 
As discussed in Section~\ref{sec:atomicity-writes}, a transaction's updates are made visible only after its commit metadata is written to storage.
Algorithm~\ref{alg:read} uses the local cache of committed transaction metadata and recent versions of keys to ensure that reads are only issued from already-committed transactions.
}

Algorithm~\ref{alg:read} shows our protocol for guaranteeing atomic reads.
If a client requests key $k$, two constraints limit the versions it can read.
First, if $k_i$ was cowritten with an earlier read $l_i$, we must return a version, $k_j$, that is at least as new as $k_i$ ($j \geq i$).
Second, if we previously read some version $l_i$, the $k_j$ return cannot be cowritten with $l_j \mid j > i$; if it was, then we should have returned $l_j$ to the client earlier, and this read set would violate read atomicity.

\begin{algorithm}[t]
\small
{
    \caption{AtomicRead: For a key $k$, return a key version $k_j$ such that the read set $R$ combined with $k_j$ does not violate Definition~\ref{def:atomic-readset}.}
    \begin{algorithmic}[1]
    \Require $k, R, WriteBuffer, storage, KeyVersionIndex$
        \State {$lower \coloneqq 0$} // Transaction ID lower bound. 
        \\ // Lines 3-5 check case~(1) of the inductive proof.
        \For {$l_i \in R$}
            \If {$k \in l_i.cowritten$} // We must read $k_j$ such that $j \geq i$.
                \State {$lower = max(lower, i)$} 
            \EndIf
        \EndFor \\
        // Get the latest version of $k$ that we are aware of.
        \State {$latest \coloneqq max(KeyVersionIndex[k])$} 
        \If {$latest == None \land lower == 0$}
            \State {return \texttt{NULL}}
        \EndIf
        \State {$target \coloneqq None$} // The version of $k$ we want to read.
        \State {$candidateVersions \coloneqq sort(filter(KeyVersionIndex[k], kv.tid \geq lower))$} // Get all versions of $k$ at least as new as $lower$.
        \\ // Loop through versions of $k$ in reverse timestamp order --- lines 13-23 check case~(2) of the inductive proof.
        \For {$t \in candidateVersions.reverse()$}
            \State{$valid \coloneqq True$} 
            \For{$l_i \in k_t.cowritten$}
                \If {$l_j \in R \land j < t$}
                    \State{$valid \coloneqq False$}
                    \State{break}
                \EndIf
            \EndFor
            \If {$valid$}
                \State {$target = t$} // The most recent valid version.
                \State {break}
            \EndIf
        \EndFor
        
        \If {$target == None$}
            \State {return \texttt{NULL}}
        \EndIf
        \State{$R_{new} \coloneqq R \cup \{k_{target}\}$}
        \State {return $storage.get(k_{target}), R_{new}$}
    \end{algorithmic}
    \label{alg:read}
}
\end{algorithm}

\begin{theorem}
\label{thm:new_atomic}
Given $k$ and $R$, the $R_{new}$ produced by Algorithm~\ref{alg:read} is an Atomic Readset, as defined in Definition~\ref{def:atomic-readset}.
\end{theorem}

\begin{proof}

We prove by induction on the size of $R$.

\textbf{Base Case}.
Before the first read is issued, $R$ is empty and is trivially an Atomic Readset.
After executing Algorithm~\ref{alg:read} for the first read, $R_{new}$ contains a single key, $k_{target}$, so Theorem~\ref{thm:new_atomic} holds; $R_{new}$ is an Atomic Readset.

\textbf{Inductive hypothesis}: 
Let $R$ be the Atomic Readset up to this point in the transaction, and let $k_{target}$ be the key version returned by Algorithm~\ref{alg:read}. 
We show that $R_{new}$ is also an Atomic Readset.
From the constraints described above, we must show that~(1) $\forall l_i \in R, k_i \in l_i.cowritten \Rightarrow target \geq i$, and~(2) $\forall l_{target} \in k_{target}.cowritten, l_i \in R \Rightarrow i \geq target$.

Lines 3-5 of Algorithm~\ref{alg:read} ensure~(1) by construction, as the lower bound of $target$ is computed by selecting the largest transaction ID in $R$ that modified $k$---we never return a version that is older than $lower$.
We iterate through versions of $k$ that are newer than $lower$, starting with the most recent version first.
Lines 13-23 check if each version satisfies case~(2)
We iterate through all the cowritten keys of each candidate version.
If any cowritten key is in $R$, we declare the candidate version valid if and only if the cowritten key's version is \emph{not newer} than the version in $R$.
If there are no valid versions, we return \texttt{NULL}.

In summary, $k_j$ satisfies case~(1) because we consider no versions older than $lower$, and it is satisfies case~(2) because we discard versions of $k$ that conflict with previous reads. \qedhere

\end{proof}

\subsection{Other Guarantees} \label{sec:atomicity-rr-ryw}

\system{} makes two other guarantees, defined in $\S$\ref{sec:atomicity-guarantees}: read-your-writes and repeatable reads. 
We briefly discuss each.

\smallitem{Read-Your-Writes}
When a transaction requests a key that is currently stored in its own write set, we simply return that data immediately; this guarantees the read-your-writes property described above.
Note that this process operates outside of the scope of Algorithm~\ref{alg:read}. 
Key versions that are stored in the \awb{} are not yet assigned a commit timestamp, so they cannot correctly participate in the algorithm.
We make this exception because read-your-writes is a useful guarantee for programmers.

\smallitem{Repeatable Read}.
Ensuring repeatable read is a corollary of Theorem~\ref{thm:new_atomic}. 
As mentioned in~\ref{sec:atomicity-guarantees}, read-your-writes takes precedence over repeatable read; therefore, Corollary~\ref{crl:rr} only applies in transactions without intervening writes of $k$.

\begin{corollary} \label{crl:rr}

Let $R$ be an Atomic Readset for a transaction $T_i$, and let $k_j, R_{new}$ be the results of Algorithm~\ref{alg:read}.
$k \notin T_i.writeset \land k_i \in R \Rightarrow k_i == k_j$.

\end{corollary}

\begin{proof}
Given $R_{new}$ and $k_j$ as the results of Algorithm~\ref{alg:read}, Theorem~\ref{thm:new_atomic} tells us that $\forall l_i \in R, k_i \in l_i.cowritten \Rightarrow j \geq i$.
Since $k_i \in R$ (and trivially, $k_i \in k_i.cowritten$), we know that $j \geq i$ for the $k_j$ returned by Algorithm~\ref{alg:read}.

Theorem~\ref{thm:new_atomic} further guarantees that $\forall l_j \in k_j.cowritten, l_i \in R \Rightarrow i \geq j$.
Once more, since $k_j \in k_j.cowritten$ and $k_i \in R$, we know that $i \geq j$.
Combining these two cases, $i == j$, meaning Algorithm~\ref{alg:read} guarantees repeatable read. \qedhere

\end{proof}

\subsection{Staleness} \label{sec:atomicity-staleness}

%


Our protocol is more flexible than RAMP because it allows interactively defined read sets.
However, it increases the potential for reading stale data because of restriction~(2) in the proof of Theorem~\ref{thm:new_atomic}.
If $T_r$ reads $l_i$, it cannot later read $k_j$ if $l_j \in k_j.cowritten, j > i$, because this violates the Definition~\ref{def:atomic-readset}.
The RAMP protocol in \cite{bailis2014ramp} avoided this pitfall with pre-declared read sets---$l$ and $k$ are read at the same time. 
If the storage engine returns the same $l_i$ and $k_j$ as above, \cite{bailis2014ramp} repairs this mismatch by forcing a read of $l_j$.
However, in our system, if $k_i$ is exposed to the client without knowledge of the later read of $l$, we cannot repair the mismatch.

In extreme cases, this can cause transaction aborts.
Continuing our example, if $T_r$ reads $l_i$ and then requests $k$, Algorithm~\ref{alg:read} will determine that $k_j$ is not a valid version for $T_r$.
If $k_j$ is the only version of $k$, we return \texttt{NULL} because $\{l_i, k_j\}$ does not form Atomic Readset.
Note that this equivalent to reading from a fixed database snapshot---if no versions of $l$ exist as of time $i$, a client would read \texttt{NULL}, abort, and retry.

\section{Scaling \system{}} \label{sec:distributed}

A key requirement for any serverless system is the ability to scale to hundreds or thousands of parallel clients.
The protocols described in $\S$\ref{sec:atomicity} ensure read atomicity for a single \system{} replica.
In this section, we discuss how we scale \system{} while maintaining distributed consistency.

As discussed earlier, we explicitly chose to avoid coordination-based techniques, as they have well-known issues with performance and scalability~\cite{brewercap, chandra2007paxos}.
\system{} nodes do not coordinate on the critical path of each transactions.
The write protocol described in $\S$\ref{sec:atomicity-writes} allows each transaction to write to separate storage locations, ensuring that different nodes do not accidentally overwrite each others' updates.

Allowing each node to commit transactions without coordination improves performance but requires ensuring nodes are aware of transactions committed by other nodes.
Each machine has a background thread that periodically---every 1 second---gathers all transactions committed recently on this node and broadcasts them to all other nodes.
This thread also listens for messages from other replicas.
When it receives a new commit set, it adds all those transactions to its local Commit Set Cache and updates its key version index.

In a distributed setting where we might be processing thousands of transactions a second, the cost of communicating this metadata can be extremely high.
We now describe an optimization that reduces \system{}'s communication overheads.
$\S$\ref{sec:distributed-ft} discusses fault-tolerance for distributed deployments.
In $\S$\ref{sec:distributed-k8s}, we describe \system{}'s deployment model.

\subsection{Pruning Commit Sets} \label{sec:distributed-pruning}

To avoid communicating unnecessary metadata, we proactively prune the set of transactions that each node multicasts.
In particular, any transaction that is locally \emph{superseded} does not need to be broadcast.
A transaction $T_i$ is locally superseded if, $\forall k_i \in T_i.writeset, \exists k_j \mid j > i$---that is, for every key written by $T_i$, there are committed versions of those keys written by transactions newer than $T_i$.
For highly contended workloads in particular---where the same data is likely to be written often---this significantly reduces the volume of metadata that must be communicated between replicas.

\begin{algorithm}[t]
\small
{
    \caption{IsTransactionSuperseded: Check whether transaction $T_i$ has been superseded---if there is a newer version of every key version written by $T_i$.}
    \begin{algorithmic}[1]
    \Require $T_i, keyVersionIndex$
        \For{$k_i \in T_i.writeset$}
            \State{$latest \coloneqq k.latest\_tid()$}
            \If {$latest == i$}
                \State{return False}
            \EndIf
        \EndFor
        
        \State{return True}
    \end{algorithmic}
    \label{alg:supersede}
}

\end{algorithm}

Algorithm~\ref{alg:supersede} how we determine if a transaction is superseded.
Each node's background multicast protocol checks whether a recently committed transaction is superseded before sending it to other replicas.
If the transaction is superseded, it is omitted entirely from the multicast message.
Similarly, for each transaction received via multicast, the receiving node checks to see if it is superseded by transactions stored locally; if it is, we do not merge it into our the metadata cache.
Note that we can safely make decisions about transaction supersedence without coordination because each transaction receives monotonically increasing sets of keys; once a transaction is superseded on a particular node, that node can safely delete the transaction metadata. 

\subsection{Fault Tolerance} \label{sec:distributed-ft}

We now turn to guaranteeing fault tolerance for the distributed protocols described in this section; we consider both safety and liveness. 
To guarantee safety, we rely on our write-ordering protocol, which ensures that each node does not persist dirty data and that commits are correctly acknowledged.
To guarantee liveness, we must further ensure that if a replica commits a transaction, acknowledges to a client, and fails before broadcasting the commit, other replicas are still made aware of the committed data.
If other nodes do not know about the committed transaction, the new data will be in storage but will never be visible to clients, which is equivalent to not having committed.

To this end, distributed deployments of \system{} have a fault manager (see Figure~\ref{fig:arch}) that lives outside of the request critical path.
The fault manager receives every node's committed transaction set without our pruning optimization applied.
It periodically scans the Transaction Commit Set in storage and checks for persisted commit records that it has not received via broadcast.
It notifies all \system{} nodes of any such transactions, ensuring that data is never lost once it has been committed.
\newtext{Thus, if a \system{} node acknowledges a commit to the client but fails before broadcasting it to other \system{} nodes, the fault manager will read that transaction's commit record and ensure that other nodes are aware of the transaction's updates.}
The fault manager is itself stateless and fault-tolerant: If it fails, it can simply scan the Commit Set again. 

\subsection{Deployment and Autoscaling} \label{sec:distributed-k8s}

\system{} is deployed using Kubernetes~\cite{kubernetes}, a cluster management tool that deploys applications running in Docker containers~\cite{docker}.
Each \system{} replica as well as the fault manager run in separate Docker containers and on separate machines.
The fault manager described above is responsible for detecting failed nodes and configuring their replacements.

An important part of any serverless system is the ability to autoscale in response to workload changes.
Our protocols for achieving distributed fault tolerance and read atomicity do not require coordination, and distributed deployments of \system{} scale with low overhead, which we will demonstrate in $\S$\ref{sec:eval-scale}.
The second aspect of this challenge is making accurate scaling decisions without user intervention.
This is a question of designing an efficient policy for choosing to add and remove nodes from the system. 
That policy is pluggable in \system{} out of scope this paper; we return to it in Section~\ref{sec:conclusion}. 
\section{Garbage Collection} \label{sec:gc}

In the protocols described thus far, there are two kinds of data that would grow monotonically if left unchecked.
The first is transaction commit metadata---the list of all transactions committed by \system{} thus far.
The second is set of key versions.
As described in $\S$\ref{sec:atomicity-writes}, each transaction's updates are written to unique keys in the storage engine and are never overwritten.
Over time, the overheads incurred from these sources can grow prohibitive, in terms of both performance and cost.
In $\S$\ref{sec:gc-local}, we describe how each node clears its local metadata cache, and in $\S$\ref{sec:gc-global}, we describe how we reduce storage overheads by deleting old data globally.

\subsection{Local Metadata Garbage Collection} \label{sec:gc-local}

In $\S$\ref{sec:distributed-ft}, we introduced Algorithm~\ref{alg:supersede}, which enables each node to locally determine whether a particular transaction has been superseded because there are newer versions of all keys the transaction wrote.
To locally garbage collect transaction metadata, a background garbage collection (GC) process periodically sweeps through all committed transactions in the metadata cache.
For each transaction, the background process executes Algorithm~\ref{alg:supersede} to check if it is superseded and ensures that no currently-executing transactions have read from that transaction's write set.

If both conditions are met, we remove that transaction from the Commit Set Cache and evict any cached data from that transaction.
This significantly reduces our metadata overheads because old transactions are frequently discarded as new transactions arrive.
While supersedence can safely be decided locally, individual nodes \emph{cannot} make decisions about whether to delete key versions because a transaction running at another node might read the superseded transaction's writes. 
As a result, we next describe a global protocol that communicates with all replicas to garbage collect key versions.
Each individual replica maintains a list of all locally deleted transaction metadata to aid in the global protocol.

\subsection{Global Data Garbage Collection} \label{sec:gc-global}

The fault manager discussed in $\S$\ref{sec:distributed} also serves as a global garbage collector (GC).
We combine these processes because the fault manager already receives commit broadcasts from \system{} nodes, which allows us to reduce communication costs.
The global GC process executes Algorithm~\ref{alg:supersede} to determine which transactions have been superseded.
It generates a list of transactions it considers superseded and asks all nodes if they have locally deleted those transactions.
If all nodes have deleted a transaction's metadata, we can be assured that no running transactions will attempt to read the deleted items.
Nodes respond with the transactions they have deleted, and when the GC process receives acknowledgements from all nodes\footnote{Note that knowing \emph{all} nodes present in the system is a traditional distributed systems membership problem, which requires coordination; we currently rely on Kubernetes to provide this information.}, it deletes the corresponding transaction's writes and commit metadata.
We allocate separate cores for the data deletion process, which allows us to batch expensive delete operations separate from the GC process.

\subsubsection{Limitation: Missing Versions} \label{sec:gc-global-limits}

There is one key shortcoming to this protocol.
Recall from Section~\ref{sec:atomicity-staleness} that \system{} returns \texttt{NULL} if there are no key versions in the valid timestamp range from Algorithm~\ref{alg:read}.
This problem can be exacerbated by our garbage collection protocol. 

As mentioned earlier, our local metadata GC protocol will not delete a transaction $T_i$ if a running transaction, $T_j$, has read from $T_i$'s write set. 
However, since we do not know each running transaction's full read set, we might delete data that would be required by a running transaction in the future.
Consider the following transactions and write sets: $T_a: \{k_a\}, T_b: \{l_b\}, T_c: \{k_c, l_c\}, a < b < c$.
Say a transaction $T_r$ first reads $k_a$ then requests key $l$. 
The GC process will not delete $T_a$ because $T_r$ has read from it and is uncommitted.
However, it might delete $T_b$ if there are no restrictions on it; when $T_r$ attempts to read $l$, it will find no valid versions since $l_c$ is invalid, and Algorithm~\ref{alg:read} will return \texttt{NULL}.

Long-running transactions accessing frequently updated keys are particularly susceptible to this pitfall.
These transactions might be forced to repeatedly retry because of missing versions, significantly increasing latencies.
In practice, we mitigate this issue by garbage collecting the oldest transactions first. 
In our evaluation, we did not encounter valid versions of keys had been deleted by the GC process.
\section{Evaluation} \label{sec:eval}

In this section, we present a detailed evaluation of \system{}.
A key design goal for \system{} is the flexibility to run on a variety of cloud storage backends, so we implemented \system{} over AWS S3, a large scale object storage system, and AWS DynamoDB, a cloud-native key-value store.
We also run \system{} over Redis (deployed via AWS ElastiCache) because it offers best-case performance for a memory-speed KVS despite the fact that it is not an autoscaling storage engine.
All experiments use Redis in cluster mode with 2 shards and 2 nodes per shard, unless stated otherwise.

First, we measure \system{}'s performance overheads and consistency benefits by comparing it to a variety of other serverless storage architectures ($\S$\ref{sec:eval-overheads}).
We then evaluate aspects of the system design: read caching ($\S$\ref{sec:eval-caching}), scalability ($\S$\ref{sec:eval-scale}), and garbage collection ($\S$\ref{sec:eval-gc}).
Finally, we measure \system{}'s ability to tolerate and recover quickly from faults ($\S$\ref{sec:eval-ft}).

\newtext{\system{} is implemented in just over 2,500 lines of Go and 700 lines of Python and runs on top of Kubernetes~\cite{kubernetes}.
The majority of the \system{} protocols described in this paper are implemented in Go.
We use Python to spin up and configure Kubernetes clusters as well as to detect node failures in Kubernetes and to reconfigure the cluster in such an event.
We use a simple stateless load balancer implemented in Go to route requests to \system{} nodes in a round-robin fashion.}
All experiments were run in the \texttt{us-east-1a} AWS availability zone (AZ).
Each \system{} node ran on a \texttt{c5.2xlarge} EC2 instance with 8vCPUs (4 physical cores) and 16GB of RAM.

\subsection{\system{} Overheads}\label{sec:eval-overheads}

We first measure the performance overheads introduced by \system{} relative to interacting with cloud storage engines without \system{}.
To isolate the overheads introduced by Function-as-a-Service platforms, we first measure IO cost with and without \system{} interposed ($\S$\ref{sec:eval-overheads-io}); we then measure end-to-end latencies and consistency anomalies for transactions running on AWS Lambda over a variety of storage engines ($\S$\ref{sec:eval-overheads-e2e}).

\subsubsection{IO Latency} \label{sec:eval-overheads-io}

\begin{figure}[t]
  \centering
    \includegraphics[width=\figwidth]{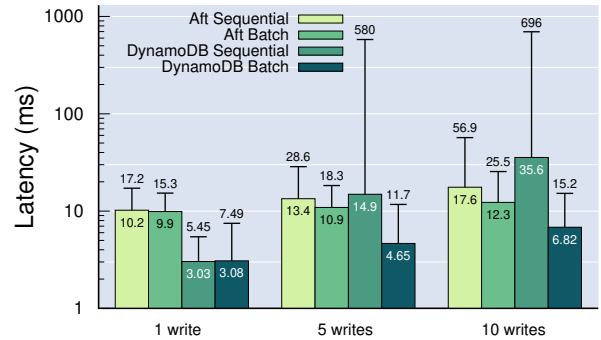}
  \caption{\small
    The median (box) and 99th percentile (whisker) latencies across 1,000 sequential requests for performing 1, 5, and 10 writes from a single client to DynamoDB and \system{} with and without batching.
    \system{}'s automatic batching allows it to significantly outperform sequential writes to DynamoDB, while its commit protocol imposes a small fixed overhead relative to batched writes to DynamoDB.
  }
  \label{fig:overhead-micro}
  \vspace{-0.5em}
\end{figure}

We first compare the cost of writing data directly to DynamoDB and to the cost of the same set of writes using \system{}'s commit protocol ($\S$\ref{sec:atomicity-writes}).
To isolate our performance overheads, we issue writes from a single thread in a VM rather than using a FaaS system.
We measure four configurations.
Our two baselines write directly to DynamoDB, one with sequential writes and the other with batching.
Batching provides best-case performance, but interactive applications can rarely batch multiple writes.
\system{} takes advantage of batched writes by default in its commit protocol---all client updates sent to the \awb{} are written to storage in a single batch when possible.
We also measure two configurations over \system{}---one where the client sends sequential writes to \system{}, and one where the client sends a single batch.

Figure~\ref{fig:overhead-micro} shows the latency for performing 1, 5, and 10 writes.
As expected, the latency of sequential writes to DynamoDB increases roughly linearly with the number of writes, and the 99th percentile latency increases super-linearly.
The latency of batched writes to DynamoDB on the other hand scales much better, increasing by about 2$\times$ from 1 write to 10.

Both \system{} configurations take advantage of batched writes.
\system{} Sequential's performance scales better than DynamoDB Sequential's, but latency increases about 70\% from 1 write to 10 writes---this is primarily due to the cost and network variance incurred by issuing sequential requests from the client to \system{}.
The \system{} Batch bar measures a client who sends all writes in a single request to \system{}, leading to much more consistent performance.
We observe a fixed difference of about 6ms between the DynamoDB Batch and the \system{} Batch measurements---this is (1) the cost of the extra network overhead imposed by shipping data to \system{}, and (2) the cost of writing the extra commit record our protocol requires.

\textit{\textbf{Takeaway}: For interactive applications that perform sequential writes,} \system{} \textit{significantly improves IO latency by automatically batching updates.} 

\subsubsection{End-to-End Latency} \label{sec:eval-overheads-e2e}

\begin{figure}[t]
  \centering
    \includegraphics[width=\figwidth]{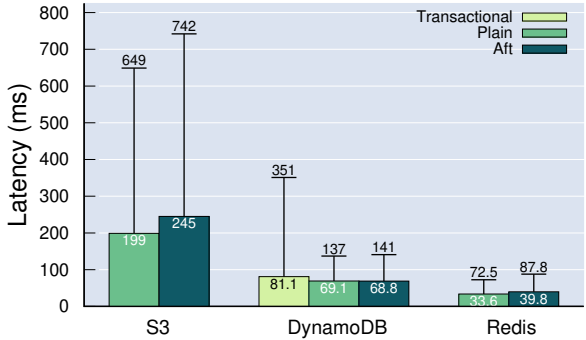}
  \caption{\small
    The end-to-end latency for executing a transaction with two sequential functions, each of which does 1 write and 2 reads (6 IOs total) on AWS S3, AWS DynamoDB, and AWS ElastiCache (Redis).
    Numbers are reported from 10 parallel clients, each running 1,000 transactions. 
  }
  \label{fig:overhead-e2e}
  \vspace{-0.5em}
\end{figure}

\begin{table}[t]
\begin{center}
\footnotesize
\begin{tabular}{|c|c|c|c|}
\hline
\textbf{\shortstack{Storage \\ Engine}} & \textbf{\shortstack{Consistency \\ Level}} & \textbf{RYW Anomalies} & \textbf{FR Anomalies} \\
\hline
\textbf{\system{}} & Read Atomic & 0 & 0 \\
\hline
\textbf{S3} & None & 595 & 836 \\
\hline
\textbf{DynamoDB} & None & 537 & 779 \\
\hline
\textbf{DynamoDB} & Serializable & 0 & 115 \\
\hline
\textbf{Redis} & \shortstack{Shard \\ Linearizable} & 215 & 383 \\
\hline

\end{tabular}
\end{center}
\caption{\small
    A count of the number of anomalies observed under Read Atomic consistency for DynamoDB, S3, and Redis over the 10,000 transactions run in Figure~\ref{fig:overhead-e2e}.
    Read-Your-Write (RYW) anomalies occur when transactions attempt to read keys they wrote and observe different versions.
    Fractured Read (FR) anomalies occurs when transactions read fractured updates with old data (see $\S$\ref{sec:bg-ra}).
    \system{}'s read atomic isolation prevents up to 13\% of transactions from observing anomalies otherwise allowed by DynamoDB and S3.
}
\label{table:overhead-inconsistencies}
\end{table}

Next, we measure the end-to-end latency of executing transactions on AWS Lambda with and without \system{} interposed between the compute and storage layers.
We evaluate three storage systems: AWS S3, AWS DynamoDB, and Redis.
Each transaction is composed of 2 functions, which perform one write and two reads each; all reads and writes are of 4KB objects.
We chose small transactions because they reflect many real-world CRUD applications and because they highlight \system{}'s overheads; we use this workload in the rest of our evaluation.
We measure performance and consistency anomalies with a lightly skewed workload (a Zipfian coefficient of 1.0).
Figure~\ref{fig:overhead-e2e} and Table~\ref{table:overhead-inconsistencies} show our results.
The bars labeled ``Plain'' in Figure~\ref{fig:overhead-e2e} represent end-to-end transaction latencies measured by running functions which write data directly to the respective storage engines.
When executing requests without \system{}, we detect consistency anomalies by embedding the same metadata \system{} uses---a timestamp, a UUID, and a cowritten key set---into the key-value pairs; this accounts for about an extra 70 bytes on top of the 4KB payload.

\smallitem{Consistency}. 
\system{}'s key advantage over DynamoDB, Redis, and S3 is its read atomic consistency guarantee.
We measure two types of anomalies here.
Read-Your-Write (RYW) anomalies occur when a transaction writes a key version and does not later read the same data; Fractured Read (FR) anomalies occur when a read violates the properties described in $\S$\ref{sec:bg-ra}---these encompass repeatable read anomalies (see $\S$\ref{sec:atomicity-rr-ryw}).
Table~\ref{table:overhead-inconsistencies} reports the number of inconsistencies observed in the 10,000 transactions from Figure~\ref{fig:overhead-e2e}.
Vanilla DynamoDB and S3 have weak consistency guarantees and incur similar numbers of anomalies---6\% of transactions experience RYW anomalies, and 8\% experience FR anomalies.

Each Redis shard is linearizable but no guarantees are made across shards.
This consistency model combined with low latency IO enables it to eliminate many anomalies by chance, as reads and writes interfere with each other less often.
Nonetheless, we still found anomalies on 6\% of requests.

We also evaluate DynamoDB's transaction mode~\cite{ddb-txn}, which provides stronger consistency than vanilla DynamoDB.
\newtext{In contrast to \system{}'s general purpose transactions, DynamoDB supports transactions that are either \emph{read-only} or \emph{write-only}.
All operations in a single transaction either succeed or fail as a group; however, DynamoDB's transaction mode \emph{does not} to guarantee atomicity for transactions that span multiple functions---each transaction is only a single API call.
To accommodate this model, we modified the workload slightly: The first function in the request does a two-read transaction, and the second function does a two-read transaction followed by a two-write transaction.
We grouped all writes into a single transaction to guarantee that the updates are installed atomically rather than being spread across two separate transactions---this reduces the flexibility of the programming model but is more favorable to DynamoDB's transactional guarantees.}

This setup avoids RYW anomalies because all of a request's writes are done in a single transaction.
However, reads are spread across two transactions in two different functions, so we still encounter FR anomalies on over 1\% of requests.
DynamoDB serializes all transactions, so many writes will be executed between the two read transactions, making our application likely to read newer data in the second read transaction that conflicts with data read in the first.

\smallitem{Performance}. 
\system{} imposes the highest overhead over S3---25\% slower than baseline at median and 14\% at 99th percentile.
S3 is a throughput-oriented object store that has high write latency variance, particularly for small objects~\cite{Brantner:2008:BDS:1376616.1376645, costlo}, and even in the \system{}-less configuration, S3 is 4 to 10$\times$ slower than other storage systems.
Our design writes each key version to a separate storage key; this is poorly suited to S3, which has high random IO latencies and is sensitive to data layouts. 
For this reason, we do not consider S3 in the rest of our evaluation.
We return to this point in $\S$\ref{sec:conclusion}.

\system{} is able to match DynamoDB Plain's performance.
In this workload, each function does one write, so when writing directly from Lambda to DynamoDB, function cannot take advantage of write batching.
On the other hand, \system{}'s use of batching offsets the cost of its extra commit metadata write.
With transaction mode enabled, DynamoDB proactively aborts transactions in the case of conflict, so the reported latencies include retries.
\system{} improves median performance over DynamoDB's transaction mode by 18\% at median and by 2.5$\times$ at the 99th percentile.

Finally, \system{} imposes a 20\% latency penalty compared to Redis Plain because we are not able to take advantage of batching.
While Redis supports a \texttt{MSET} operation to write multiple keys at once, that operation can only modify keys in a single shard.
Since requests can write arbitrary data, we are not guaranteed to modify keys in a single shard and cannot consistently batch updates.

\textit{\textbf{Takeaway}:} \system{} \textit{offers performance that is competitive with state-of-the-art cloud storage engines while also eliminating a significant number of anomalies.}

\subsection{Read Caching \& Data Skew} \label{sec:eval-caching}
\begin{figure}[t]
  \centering
    \includegraphics[width=\figwidth]{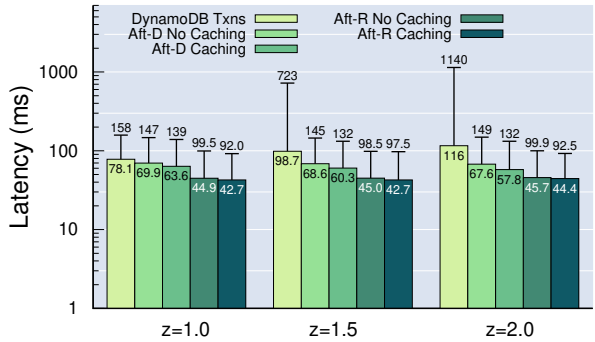}
  \caption{\small
    End-to-end latency for \system{} over DynamoDB (\system{}-D) and Redis (\system{}-R) with and without read caching enabled, as well as DynamoDB's transaction mode. 
    We vary the skew of the data access distribution to demonstrate the effects of contended workloads.
    Caching improves \system{}-D's performance by up to 15\%, while it has little effect on \system{}-R's performance.
    DynamoDB's transaction mode suffers under high contention due to large numbers of repeated retries.
  }
  \label{fig:caching}
  \vspace{-0.5em}
\end{figure} 

We now turn our attention to \system{}'s data caching, its effect on performance, and its interaction with access distribution skewness.
In this experiment, we use the same workload as in Section~\ref{sec:eval-overheads-e2e}---a 2-function transaction with 2 reads and 1 write per function.
Figure~\ref{fig:caching} shows the median and 99th percentile latencies for \system{} deployed over DynamoDB (\system{}-D) and Redis (\system{}-R) with and without caching.
We also measure DynamoDB's transaction mode; we omit any configurations that do not provide transactions in some form.
We measured 3 Zipfian distributions: 1.0 (lightly contended), 1.5 (moderately contended), and 2.0 (heavily contended).

Interestingly, we find that \system{}-R's performance varies very little across all configurations.
Read caching does not improve performance because the cost of fetching a 4KB payload from Redis is negligible compared to invoking a Lambda function and executing our read and write protocols.
While recent work has shown that Redis' performance can suffer under high contention~\cite{wu2019anna}, this experiment measures latencies and thus does not saturate Redis's capacity.

With caching, \system{}-D's performance improves by 10\% for the lightly contended workload and up to 17\% for the heavily contended workload.
As distribution skew increases, so does the likelihood that we have a valid key version cached, thus improving performance.

Finally, we measure DynamoDB's transaction mode.
Interestingly, we find that for Zipf=1.0, DynamoDB's transactional performance improves relative to Figure~\ref{fig:overhead-e2e} in $\S$\ref{sec:eval-overheads-e2e}.
This is because we use a larger dataset in this experiment (100,000 keys vs 1,000 keys), so the lightly contended workload is less likely to encounter data access conflicts.
As we increase contention, performance degrades significantly, and for the most contended workload, \system{}-D is 2$\times$ faster at median and 7.6$\times$ better at the tail.

\textit{\textbf{Takeaway}: Introducing read caching unlocks significant performance improvements for} \system{}, \textit{particularly for skewed access distributions.}

\subsection{Read-Write Ratios} \label{sec:eval-rw-ratio}

\newtext{
Next, we look at the effects of read-write ratios within a single transaction.
Thus far, our workload has consisted of transactions with only 4 writes and 2 reads split across 2 functions.
In this section, we will use longer transactions with 10 total IOs and vary the percentage of those IOs that are reads from 0\% to 100\%.
}

\begin{figure}[t]
  \centering
    \includegraphics[width=\figwidth]{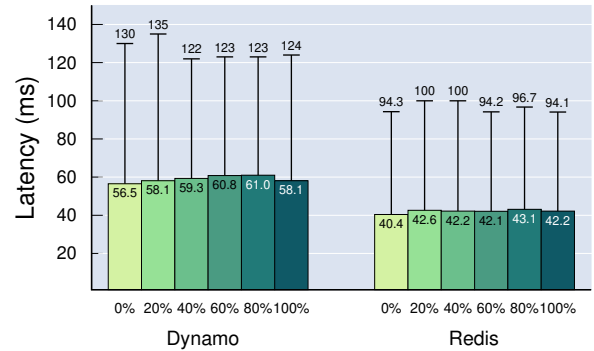}
  \caption{\small
    Median and 99th percentile latency for \system{} over DynamoDB and Redis as a function of read-write ratio, from transactions with 0\% reads to transactions with 100\% reads.
    \system{} over Redis should little variation, while our use of batching over DynamoDB leads to small effects based on read-write ratios.
  }
  \label{fig:rw-ratio}
  \vspace{-0.5em}
\end{figure}

\newtext{
Figure~\ref{fig:rw-ratio} shows our results.
With \system{} running over DynamoDB, performance is largely consistent with less than 10\% variance.
There is a slight increase in median latency from 0\% reads to 80\% reads.
At 0\% reads, there are two API calls to DynamoDB---one to write the batch of updates, and the other to write the transaction's commit record.
As we add reads, each individual read results in a separate API call, resulting in a slight increase in latency.
At 100\% reads, we remove the batch write API call, leading to a small dip in latency.
0\% and 20\% reads (i.e., 10 and 8 writes, respectively) have higher tail latencies because the larger numbers of writes increase the chance that the batch writes hit slower shards in the underlying storage engine---this is similar to the differences in tail latencies seen in Figure~\ref{fig:overhead-micro}.
}

\newtext{
\system{} over Redis shows very little variation in performance across all read-write ratios.
This is because Redis (in cluster mode) does not support writing multiple objects at once, and the system treats reads and writes roughly uniformly.
As a result, for all configurations, we make 11 API calls---10 for the IOs and 1 for the final commit record.
}

\newtext{
\textit{\textbf{Takeaway}:} \system{} \textit{maintains consistent performance across a variety of read-write ratios with minor variations based on the characteristics of the underlying storage system.}
}

\subsection{Transaction Length} \label{sec:eval-txn-length}

\begin{figure}[t]
  \centering
    \includegraphics[width=\figwidth]{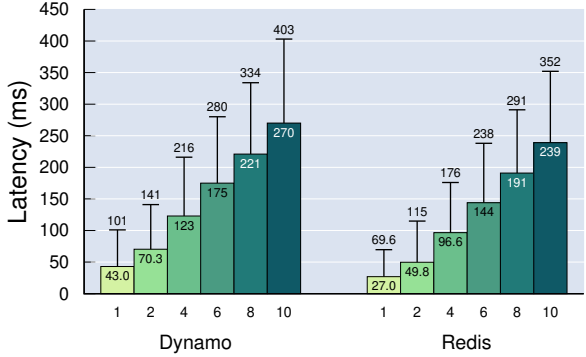}
  \caption{\small
    Median and 99th percentile latency for \system{} over DynamoDB and Redis as a function of transaction length, from 1 function (3 IOs) to 10 functions (30 IOs).
    Longer transactions mask the overheads of \system{}'s protocols, which play a bigger role in the performance of the shorter transactions.
  }
  \label{fig:txn-length}
  \vspace{-0.5em}
\end{figure}

\newtext{In this experiment, we study the performance impact of transaction length. 
We vary transactions from 1 function to 10 functions, where each function consists of 2 reads and 1 write.
Figure~\ref{fig:txn-length} shows our results.}

\newtext{
\system{} scales roughly linearly with transaction length over both DynamoDB and Redis.
As in previous experiments, \system{}'s use of DynamoDB's batching capabilities means that the overhead of increased writes is masked with the batch update operation.
As a result, 10-function transactions are only 6.2$\times$ slower than 1-function transactions---intuitively, this corresponds to the fact that $\frac23$ (or 67\%) of our API calls are reads, which scale linearly, while the remaining writes are batches into one API call.
}

\newtext{
As before, Redis requires separate API calls for each write operation and thus 10-function transactions are 8.9$\times$ slower than 1-function transactions.
The remaining difference is because the cost of writing an extra commit record to storage as a fixed cost is a large portion of the operating time of a 1-function transaction but a much smaller portion for 10-function transactions.
DynamoDB is 59\% slower than Redis for 1-function transactions but only 13\% slower for 10-function transactions.
In the rest of our experiments, we use 2-function transactions because they more clearly highlight our overheads than longer transactions do.
}

\newtext{
\textit{\textbf{Takeaway}:} \system{} \textit{scales linearly with transaction length and is able to mask update and commit overheads for longer transactions.}
}

\subsection{Scalability} \label{sec:eval-scale}

In this section, we evaluate \system{}'s scalability relative to the number of parallel clients.
We first measure the number of clients a single \system{} node can support in $\S$\ref{sec:eval-scale-single}.
We then measure the overhead of the distributed protocols in $\S$\ref{sec:distributed} by measuring throughput in distributed deployments.

\subsubsection{Single-Node Scalability} \label{sec:eval-scale-single}

\begin{figure}[t]
  \centering
    \includegraphics[width=\figwidth]{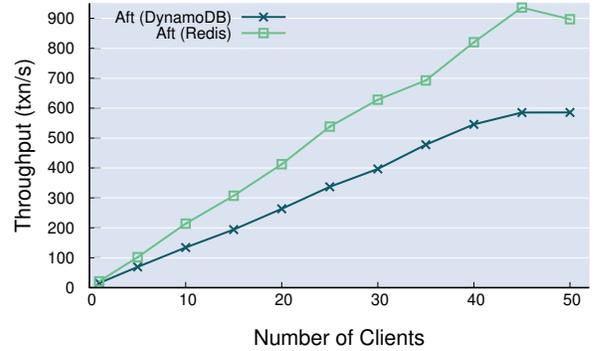}
  \caption{\small
    The throughput of a single \system{} node as a function of number of simultaneous clients issues requests to it.
    We can see a single node scales linearly until about 40 clients for DynamoDB and 45 clients for Redis, at which point, the throughput plateaus.
  }
  \label{fig:scale-single}
  \vspace{-0.5em}
\end{figure}

In this section, we measure the number of parallel clients that a single \system{} node can support.
We run the same 2-function, 6-IO transactions as before under the moderate contention level (Zipf=1.5), and we we vary the number of clients from 1 to 50.
Each client makes 1,000 requests by synchronously invoking the transaction, waiting for a response, and then triggering another transaction.
Figure~\ref{fig:scale-single} shows our results with \system{} deployed over DynamoDB and Redis.

We find  that \system{} scales linearly until 40 and 45 clients for DynamoDB and Redis, respectively.
At this point, contention for shared data structures causes \system{}'s throughput to plateau.
Similar to previous experiments, \system{} over Redis achieves better performance than \system{} over DynamoDB.
Because Redis offers significantly lower IO latencies, each transaction completes faster (see Figure~\ref{fig:overhead-e2e}).
Our clients synchronously invoke each Lambda, so the reduced latency directly translates to better throughput.
At peak, \system{} over Redis is able to achieve 900 transaction per second, while \system{} over DynamoDB achieves just under 600 transaction per second.

\textit{\textbf{Takeaway}: A single} \system{} \textit{node is able to scale linearly to over 40 clients (600 tps), demonstrating the low overhead of our read atomic protocols.}

\subsubsection{Distributed Scalability} \label{sec:eval-scale-multi}

\begin{figure}[t]
  \centering
    \includegraphics[width=\figwidth]{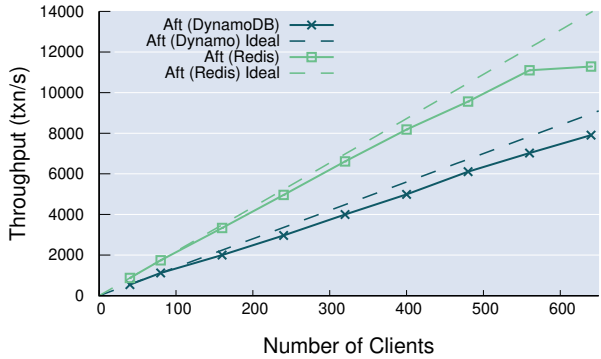}
  \caption{\small
    \system{} is able to smoothly scale to hundreds of parallel clients and thousands of transactions per second while deployed over both DynamoDB and Redis.
    We saturate either DynamoDB's throughput limits or AWS Lambda's concurrent function inovcation limit while scaling within 90\% of ideal throughput.
  }
  \label{fig:scale-multi}
  \vspace{-0.5em}
\end{figure}

We now turn to measuring \system{}'s ability to scale smoothly for multi-node deployments.
Based on the results in $\S$\ref{sec:eval-scale-single}, we ran 40 clients per \system{} node and progressively scaled up the number of nodes.
Figure~\ref{fig:scale-multi} shows our results.

Deployed over DynamoDB, we find that \system{} is able to scale seamlessly to 8,000 transactions per second for 640 parallel clients.
\system{} scales at slope that is within 90\% of the ideal slope, where the ideal throughput is the number of nodes multiplied by a single node's throughput.
We originally intended to demonstrate scalability to a thousand parallel clients, but we were restricted by AWS DynamoDB's resource limits, which would not let us scale beyond what is shown. 

With Redis, we observe that \system{} is similarly able to scale linearly.
Similar to $\S$\ref{sec:eval-scale-single}, \system{} over Redis has a higher aggregate throughput due to lower IO latencies.
Nonetheless, throughput remains within 90\% of ideal.
Throughput for 640 clients plateaus not because of \system{} overheads but because we were limited by the number of concurrent function invocations supported by AWS Lambda.

Note that we manually configured both storage systems with the appropriate resources.
We chose to disable DynamoDB's autoscaling because the goal of this experiment was not to measure efficacy of their autoscaling policy.
In general, however, DynamoDB's autoscaling support makes it well-suited to serverless applications, while Redis is a fixed-deployment system with high reconfiguration overheads.

\textit{\textbf{Takeaway}:} \system{} \textit{is able to efficiently scale to thousands of transactions per second and hundreds of parallel clients within 90\% of ideal throughput.}

\subsection{Garbage Collection Overheads} \label{sec:eval-gc}

\begin{figure}[t]
  \centering
    \includegraphics[width=\figwidth]{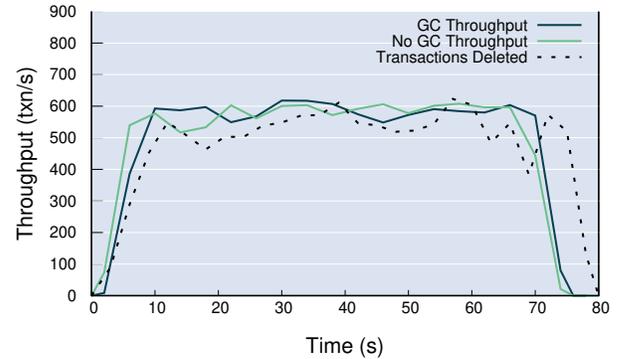}
  \caption{\small
    Throughput for \system{} over DynamoDB with and without global data garbage collection enabled.
    The garbage collection process has no effect on throughput while effectively deleting transactions at the same rate \system{} processes them under a moderately contended workload (Zipf=1.5).
  }
  \label{fig:gc}
  \vspace{-0.5em}
\end{figure}

In this experiment, we quantify the overhead of enabling \system{}'s global garbage collection ($\S$\ref{sec:gc-global}).
We run a single \system{} node with 40 clients and measure throughput with garbage collection enabled and disabled. 
We also measured the number of transactions deleted per second when garbage collection is enabled.
Figure~\ref{fig:gc} shows our results.

There is no discernible difference between throughput with garbage collection enabled and disabled.
The bulk of the work related to determining transaction supersedence happens periodically on each node to reduce metadata overheads (see $\S$\ref{sec:distributed-pruning}).
As a result, the global GC protocol simply collects lists of superseded transactions from all \system{} nodes and deletes transactions that all nodes consider superseded.

The cost of this garbage collection process is that we require separate cores that are dedicated to deleting old data.
However, the resources allocated to garbage collection are much smaller than the resources required to run the system: For the four cores we used to run the \system{} node, we only required 1 core to delete transactions.

\textbf{\textit{Takeaway}:} \system{}'s \textit{local metadata garbage collection enables efficient global deletion of superseded data with no effect on system throughput and at reasonable added cost.}

\subsection{Fault Tolerance} \label{sec:eval-ft}

\begin{figure}[t]
  \centering
    \includegraphics[width=\figwidth]{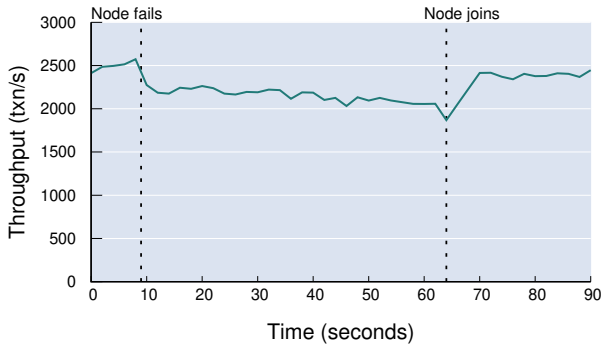}
  \caption{\small
    \system{}'s fault manager is able to detect faults and allocate new resources within a reasonable time frame; the primary overheads we observe are due to the cost of downloading Docker containers and warming up \system{}'s metadata cache.
    \system{}'s performance does not suffer significantly in the interim.
  }
  \label{fig:ft}
  \vspace{-0.5em}
\end{figure}

Finally, in this experiment, we measure \system{}'s performance in the presence of failures.
We know from our protocols and prior experiments that the behavior and performance of \system{} nodes is independent. 
In this experiment, we are looking to measure the effect of a node failure and the cost of recovering from failure---that is how long it takes for a node to cold start and join the system.
Figure~\ref{fig:ft} shows our results.

We run \system{} with 4 nodes and 200 parallel clients and terminate a node just before 10 seconds. 
We see that throughput immediately drops about 16\%.
Within 5 seconds, the \system{} management process determines that a node has failed, and it adds assigns a new node to join the cluster.
Note that we pre-allocate ``standby'' nodes to avoid having to wait for new EC2 VMs to start, which can take up to 3 minutes.

Over the next 45 seconds, this new node downloads the \system{} Docker container and updates its local metadata cache.
Just after 60 seconds, the node joins the cluster, and throughput returns to pre-failure peaks within a few seconds. 
Note that throughput is on a slight downward slope between 10 and 60 seconds---this is because the remaining three nodes are saturated, and the queue of pending requests grows, causing throughput to decrease.

Note that the overheads involved in starting a new node can be further mitigated by downloading containers in advance and by maintaining standbys nodes with warm metadata caches. 
These are engineering tasks that are not fundamental to the design of \system{}.

\textbf{\textit{Takeaway}:} \system{}'s \textit{fault management system is able to detect and recover from faults in a timely fashion.}

\section{Related Work} \label{sec:related}

\smallitem{Atomic Reads}.
The RAMP protocols in~\cite{bailis2014ramp} are the only prior work we are aware of that explicitly addresses read atomic isolation.
However, another coordination-free consistency model that guarantees atomic reads atomic is transactional causal consistency (TCC), implemented in systems like Occult~\cite{akbar2017icant} and Cure~\cite{akkoorath2016cure}.
In addition to atomic reads, causal consistency guarantees that reads and writes respect Lamport's ``happens-before'' relation~\cite{lamport1998part}.
In general, causal systems must track each transaction's read set in addition to its write set, which adds a metadata overhead that read atomicity does not require.

Two key aspects distinguish our approach from prior work.
First, \cite{bailis2014ramp}, \cite{akbar2017icant}, and \cite{akkoorath2016cure} achieve read atomicity at the storage layer, whereas \system{} is a shim above the storage engine.
This allows for more flexibility as \system{} users can pick any key-value storage system, while still maintaining consistency.
Second, the mechanisms used to achieve read atomic isolation in \cite{akbar2017icant} and \cite{akkoorath2016cure} rely on fixed node membership at the storage layer, which we cannot assume in an autoscaling serverless environment.
The read atomic isolation protocols in this paper do not require knowledge of node membership.
\system{} only requires membership for garbage collection of data, which happens off the critical path of transaction processing.

\smallitem{Multi Versioning}.
The idea of multi-version concurrency control and storage dates back to the 1970s~\cite{reed1978naming,bernsteingoodman}. 
The original Postgres storage manager~\cite{stonebraker1987pgsql} introduced the notion of maintaining a commit log separate from data version storage, an idea we adopted in \system{}.
More recently, a variety of systems have proposed and evaluated techniques for multi-version concurrency \cite{Wu:2017:EEI:3067421.3067427,Neumann:2015:FSM:2723372.2749436,Faleiro:2015:RSM:2809974.2809981}.
These systems all offer some form of strong transactional consistency (e.g., snapshot isolation, serializability), which \system{} and the RAMP protocols do not.
Similar to the causal approaches, these systems also enforce consistency in the storage system, while \system{} offers consistency over a variety of storage backends.

\smallitem{Fault Tolerance}.
There is a rich literature on fault tolerance for distributed systems (see, e.g.,~\cite{randell1975system, vstbook}).
Many techniques are conceivably applicable in the FaaS context, from checkpoint/restart of containers or virtual machines (e.g.~\cite{li2015comparing}) to log- (or ``lineage-'') based replay (e.g.~\cite{wang2019lineage}) to classical process pairs (e.g.,~\cite{bartlett1987fault}).
Our approach is based on the simple retry-from-scratch model used by existing serverless platforms such as AWS Lambda and Google Cloud Functions, and appropriate to short-lived interactive (``transactional'') tasks. 
As noted above, existing FaaS platforms attempt to offer at-least-once execution semantics, but may expose fractional writes from failed function attempts. 
\system{}'s built-in atomicity and idempotence guarantees allow a simple retry scheme to achieve exact-once semantics that provides both safety and liveness in the face of failures.
\section{Conclusion and Future Work} \label{sec:conclusion}

In this paper, we presented \system{}, a low-overhead fault-tolerance shim for serverless computing.
\system{} interposes between commodity FaaS platforms and key-value stores to achieve fault-tolerance by transparently guaranteeing read atomic isolation~\cite{bailis2014ramp}.
We develop new distributed protocols for read atomicity, which build on shared storage for high throughput and do not require pre-declared read and write sets.
\system{} adds minimal overhead to existing serverless architectures and scales linearly with the size of the cluster, while offering exactly-once execution in the face of failures.

While \system{} brings fault tolerance guarantees to FaaS programmers for the first time, we believe there are a variety of interesting avenues for future work.

\smallitem{Efficient Data Layout}.
As observed in $\S$\ref{sec:eval-overheads-e2e}, \system{}'s key-per-version data layout works well over DynamoDB and Redis, which are optimized for small key access, but performs poorly over S3.
In settings with high-volume updates, we would like to optimize data layouts for a system like S3.
\system{}'s design allows us to explore techniques for high-volume, write heavy workloads, inspired by log-structured merge trees~\cite{o1996log} and log-structured file systems~\cite{rosenblum1992design}, where writes are appended as efficiently as possible, and re-organized offline into read-optimized data structures.



\smallitem{Autoscaling Policies}.
As a shim layer, \system{} allows us to autoscale fault-tolerance infrastructure independently of both storage and compute.
Policies for efficient and accurate autoscaling decisions are a natural next step.
There are two aspects to this problem.
From a functionality perspective, we must accurately measure system load per node and effectively manage cluster size to meet application requirements.
In terms of scheduling, economic models and SLA design may offer a useful angle for engineering the interplay of potentially unbounded platform costs and user behavior.



\smallitem{Data and Cache Partitioning}.
Partitioned caching is a well-studied technique in large-scale web systems~\cite{breslau1999web}.
Na\"ive caching schemes like \system{}'s will result in every node caching largely the same data, particularly for skewed workloads.
It may be preferable to intelligently partition the space of keys across nodes~\cite{breslau1999web} to make better use of global memory while maintaining load balance. 
Doing this in \system{} would mean that each transaction would interact with multiple \system{} nodes, raising interesting new design challenges for maintaining read atomic consistency.

\smallitem{Recovery for Long-Running FaaS Programs}
As discussed in $\S$\ref{sec:related}, our retry approach could be augmented with recovery-based techniques such as checkpointing and logging/lineage, so that long-running FaaS programs need not restart from scratch. 
These mechanisms are not inconsistent with our atomicity design, but details need to be worked out---e.g., to ensure that metadata management and garbage are handled correctly upon recovery.



\smallitembot

The design of \system{} as a disaggregated shim layer enables us to study these topics in the context of commodity serverless infrastructure--even as the computing and storage platforms evolve.

\bibliographystyle{plain}
\bibliography{references}

\begin{thebibliography}{10}

\bibitem{akkoorath2016cure}
Deepthi~Devaki Akkoorath, Alejandro~Z Tomsic, Manuel Bravo, Zhongmiao Li, Tyler
  Crain, Annette Bieniusa, Nuno Pregui{\c{c}}a, and Marc Shapiro.
\newblock Cure: Strong semantics meets high availability and low latency.
\newblock In {\em 2016 IEEE 36th International Conference on Distributed
  Computing Systems (ICDCS)}, pages 405--414. IEEE, 2016.

\bibitem{akkus2018sand}
Istemi~Ekin Akkus, Ruichuan Chen, Ivica Rimac, Manuel Stein, Klaus Satzke,
  Andre Beck, Paarijaat Aditya, and Volker Hilt.
\newblock {SAND}: Towards high-performance serverless computing.
\newblock In {\em 2018 USENIX Annual Technical Conference (USENIX ATC 18)},
  pages 923--935, 2018.

\bibitem{awscasestudies}
Aws {L}ambda - case studies.
\newblock \url{https://aws.amazon.com/lambda/resources/customer-case-studies/}.

\bibitem{bailis2014ramp}
Peter Bailis, Alan Fekete, Joseph~M. Hellerstein, Ali Ghodsi, and Ion Stoica.
\newblock Scalable atomic visibility with ramp transactions.
\newblock In {\em Proceedings of the 2014 ACM SIGMOD International Conference
  on Management of Data}, SIGMOD '14, pages 27--38, New York, NY, USA, 2014.
  ACM.

\bibitem{baldini2017serverless}
Ioana Baldini, Paul Castro, Kerry Chang, Perry Cheng, Stephen Fink, Vatche
  Ishakian, Nick Mitchell, Vinod Muthusamy, Rodric Rabbah, Aleksander
  Slominski, et~al.
\newblock Serverless computing: Current trends and open problems.
\newblock In {\em Research Advances in Cloud Computing}, pages 1--20. Springer,
  2017.

\bibitem{bartlett1987fault}
Joel Bartlett, Jim Gray, and Bob Horst.
\newblock Fault tolerance in tandem computer systems.
\newblock In {\em The Evolution of Fault-Tolerant Computing}, pages 55--76.
  Springer, 1987.

\bibitem{beame2014skew}
Paul Beame, Paraschos Koutris, and Dan Suciu.
\newblock Skew in parallel query processing.
\newblock In {\em Proceedings of the 33rd ACM SIGMOD-SIGACT-SIGART symposium on
  Principles of database systems}, pages 212--223. ACM, 2014.

\bibitem{bernsteingoodman}
Philip~A. Bernstein and Nathan Goodman.
\newblock Multiversion concurrency control\&mdash;theory and algorithms.
\newblock {\em ACM Trans. Database Syst.}, 8(4):465--483, December 1983.

\bibitem{Brantner:2008:BDS:1376616.1376645}
Matthias Brantner, Daniela Florescu, David Graf, Donald Kossmann, and Tim
  Kraska.
\newblock Building a database on s3.
\newblock In {\em Proceedings of the 2008 ACM SIGMOD International Conference
  on Management of Data}, SIGMOD '08, pages 251--264, New York, NY, USA, 2008.
  ACM.

\bibitem{breslau1999web}
Lee Breslau, Pei Cao, Li~Fan, Graham Phillips, Scott Shenker, et~al.
\newblock Web caching and zipf-like distributions: Evidence and implications.
\newblock In {\em Ieee Infocom}, volume~1, pages 126--134. INSTITUTE OF
  ELECTRICAL ENGINEERS INC (IEEE), 1999.

\bibitem{brewercap}
E.~{Brewer}.
\newblock Cap twelve years later: How the ``rules'' have changed.
\newblock {\em Computer}, 45(2):23--29, Feb 2012.

\bibitem{chandra2007paxos}
Tushar~D Chandra, Robert Griesemer, and Joshua Redstone.
\newblock Paxos made live: an engineering perspective.
\newblock In {\em Proceedings of the twenty-sixth annual ACM symposium on
  Principles of distributed computing}, pages 398--407. ACM, 2007.

\bibitem{ddb-txn}
Amazon dynamodb transactions: How it works - amazon dynamodb.
\newblock
  \url{https://docs.aws.amazon.com/amazondynamodb/latest/developerguide/transaction-apis.html}.

\bibitem{docker}
Enterprise application container platform | docker.
\newblock \url{https://www.docker.com}.

\bibitem{Faleiro:2015:RSM:2809974.2809981}
Jose~M. Faleiro and Daniel~J. Abadi.
\newblock Rethinking serializable multiversion concurrency control.
\newblock {\em Proc. VLDB Endow.}, 8(11):1190--1201, July 2015.

\bibitem{fouladi2019laptop}
Sadjad Fouladi, Francisco Romero, Dan Iter, Qian Li, Shuvo Chatterjee, Christos
  Kozyrakis, Matei Zaharia, and Keith Winstein.
\newblock From laptop to {L}ambda: Outsourcing everyday jobs to thousands of
  transient functional containers.
\newblock In {\em 2019 USENIX Annual Technical Conference (USENIX ATC 19)},
  pages 475--488, 2019.

\bibitem{excamera}
Sadjad Fouladi, Riad~S. Wahby, Brennan Shacklett, Karthikeyan~Vasuki
  Balasubramaniam, William Zeng, Rahul Bhalerao, Anirudh Sivaraman, George
  Porter, and Keith Winstein.
\newblock Encoding, fast and slow: Low-latency video processing using thousands
  of tiny threads.
\newblock In {\em 14th {USENIX} Symposium on Networked Systems Design and
  Implementation ({NSDI} 17)}, pages 363--376, Boston, MA, 2017. {USENIX}
  Association.

\bibitem{gan2019open}
Yu~Gan, Yanqi Zhang, Dailun Cheng, Ankitha Shetty, Priyal Rathi, Nayan Katarki,
  Ariana Bruno, Justin Hu, Brian Ritchken, Brendon Jackson, et~al.
\newblock An open-source benchmark suite for microservices and their
  hardware-software implications for cloud \& edge systems.
\newblock In {\em Proceedings of the Twenty-Fourth International Conference on
  Architectural Support for Programming Languages and Operating Systems}, pages
  3--18. ACM, 2019.

\bibitem{serverless-cidr19}
Joseph~M. Hellerstein, Jose~M. Faleiro, Joseph Gonzalez, Johann
  Schleier{-}Smith, Vikram Sreekanti, Alexey Tumanov, and Chenggang Wu.
\newblock Serverless computing: One step forward, two steps back.
\newblock In {\em {CIDR} 2019, 9th Biennial Conference on Innovative Data
  Systems Research, Asilomar, CA, USA, January 13-16, 2019, Online
  Proceedings}, 2019.

\bibitem{bv-serverless}
Eric Jonas, Johann Schleier-Smith, Vikram Sreekanti, Chia-Che Tsai, Anurag
  Khandelwal, Qifan Pu, Vaishaal Shankar, Joao Menezes~Carreira, Karl Krauth,
  Neeraja Yadwadkar, Joseph Gonzalez, Raluca~Ada Popa, Ion Stoica, and David~A.
  Patterson.
\newblock Cloud programming simplified: A {B}erkeley view on serverless
  computing.
\newblock Technical Report UCB/EECS-2019-3, EECS Department, University of
  California, Berkeley, Feb 2019.

\bibitem{pywren}
Eric Jonas, Shivaram Venkataraman, Ion Stoica, and Benjamin Recht.
\newblock Occupy the cloud: Distributed computing for the 99{\%}.
\newblock {\em CoRR}, abs/1702.04024, 2017.

\bibitem{pocket}
Ana Klimovic, Yawen Wang, Patrick Stuedi, Animesh Trivedi, Jonas Pfefferle, and
  Christos Kozyrakis.
\newblock Pocket: Elastic ephemeral storage for serverless analytics.
\newblock In {\em 13th $\{$USENIX$\}$ Symposium on Operating Systems Design and
  Implementation ($\{$OSDI$\}$ 18)}, pages 427--444, 2018.

\bibitem{kubernetes}
Kubernetes: Production-grade container orchestration.
\newblock \url{http://kubernetes.io}.

\bibitem{lambda-best-practices}
Make a lambda function idempotent.
\newblock
  \url{https://aws.amazon.com/premiumsupport/knowledge-center/lambda-function-idempotent/}.

\bibitem{lamport1998part}
Leslie Lamport.
\newblock The part-time parliament.
\newblock {\em ACM Transactions on Computer Systems (TOCS)}, 16(2):133--169,
  1998.

\bibitem{li2015comparing}
Wubin Li and Ali Kanso.
\newblock Comparing containers versus virtual machines for achieving high
  availability.
\newblock In {\em 2015 IEEE International Conference on Cloud Engineering},
  pages 353--358. IEEE, 2015.

\bibitem{akbar2017icant}
Syed~Akbar Mehdi, Cody Littley, Natacha Crooks, Lorenzo Alvisi, Nathan Bronson,
  and Wyatt Lloyd.
\newblock I can{\textquoteright}t believe it{\textquoteright}s not causal!
  scalable causal consistency with no slowdown cascades.
\newblock In {\em 14th {USENIX} Symposium on Networked Systems Design and
  Implementation ({NSDI} 17)}, pages 453--468, Boston, MA, March 2017. {USENIX}
  Association.

\bibitem{Neumann:2015:FSM:2723372.2749436}
Thomas Neumann, Tobias M\"{u}hlbauer, and Alfons Kemper.
\newblock Fast serializable multi-version concurrency control for main-memory
  database systems.
\newblock In {\em Proceedings of the 2015 ACM SIGMOD International Conference
  on Management of Data}, SIGMOD '15, pages 677--689, New York, NY, USA, 2015.
  ACM.

\bibitem{o1996log}
Patrick O’Neil, Edward Cheng, Dieter Gawlick, and Elizabeth O’Neil.
\newblock The log-structured merge-tree (lsm-tree).
\newblock {\em Acta Informatica}, 33(4):351--385, 1996.

\bibitem{randell1975system}
Brian Randell.
\newblock System structure for software fault tolerance.
\newblock {\em Ieee transactions on software engineering}, pages 220--232,
  1975.

\bibitem{reed1978naming}
David~Patrick Reed.
\newblock {\em Naming and synchronization in a decentralized computer system.}
\newblock PhD thesis, Massachusetts Institute of Technology, 1978.

\bibitem{rosenblum1992design}
Mendel Rosenblum and John~K Ousterhout.
\newblock The design and implementation of a log-structured file system.
\newblock {\em ACM Transactions on Computer Systems (TOCS)}, 10(1):26--52,
  1992.

\bibitem{stonebraker1987pgsql}
Michael Stonebraker.
\newblock The design of the postgres storage system.
\newblock In {\em Proceedings of the 13th International Conference on Very
  Large Data Bases}, VLDB '87, pages 289--300, San Francisco, CA, USA, 1987.
  Morgan Kaufmann Publishers Inc.

\bibitem{Taft:2014:EFE:2735508.2735514}
Rebecca Taft, Essam Mansour, Marco Serafini, Jennie Duggan, Aaron~J. Elmore,
  Ashraf Aboulnaga, Andrew Pavlo, and Michael Stonebraker.
\newblock E-store: Fine-grained elastic partitioning for distributed
  transaction processing systems.
\newblock {\em Proc. VLDB Endow.}, 8(3):245--256, November 2014.

\bibitem{van2017spec}
Erwin Van~Eyk, Alexandru Iosup, Simon Seif, and Markus Th{\"o}mmes.
\newblock The {SPEC} cloud group's research vision on {FaaS} and serverless
  architectures.
\newblock In {\em Proceedings of the 2nd International Workshop on Serverless
  Computing}, pages 1--4. ACM, 2017.

\bibitem{vstbook}
Maarten van Steep and Andrew~S. Tanenbaum.
\newblock {\em Distributed Systems}.
\newblock CreateSpace independen Publishing Platform, 3.01 edition, 2018.

\bibitem{wang2019lineage}
Stephanie Wang, John Liagouris, Robert Nishihara, Philipp Moritz, Ujval Misra,
  Alexey Tumanov, and Ion Stoica.
\newblock Lineage stash: fault tolerance off the critical path.
\newblock In {\em Proceedings of the 27th ACM Symposium on Operating Systems
  Principles}, pages 338--352. ACM, 2019.

\bibitem{wu2019anna}
Chenggang Wu, Jose Faleiro, Yihan Lin, and Joseph Hellerstein.
\newblock Anna: A kvs for any scale.
\newblock {\em IEEE Transactions on Knowledge and Data Engineering}, 2019.

\bibitem{Wu:2017:EEI:3067421.3067427}
Yingjun Wu, Joy Arulraj, Jiexi Lin, Ran Xian, and Andrew Pavlo.
\newblock An empirical evaluation of in-memory multi-version concurrency
  control.
\newblock {\em Proc. VLDB Endow.}, 10(7):781--792, March 2017.

\bibitem{costlo}
Zhe Wu, Curtis Yu, and Harsha~V. Madhyastha.
\newblock Costlo: Cost-effective redundancy for lower latency variance on cloud
  storage services.
\newblock In {\em 12th {USENIX} Symposium on Networked Systems Design and
  Implementation ({NSDI} 15)}, pages 543--557, Oakland, CA, May 2015. {USENIX}
  Association.

\end{thebibliography}

\end{document}